# Thermoregulation Variation in Vertebrates Reveals Differences in Thermal Fatigue Resistance of Bones

Parker R. Brewster,[1] Jake E. Akins,[1] Casey M. Holycross[2], and Farhad Farzbod[1]

[1] *Department of Mechanical Engineering, University of Mississippi, University, MS 38677, USA*

[2] *Aerospace Systems Directorate (AFRL/RQTI), Wright-Patterson AFB, OH 45433, USA*

**Abstract**

In this study, we propose the hypothesis that there is a significant difference in thermal cycling fatigue resistance between the bones of ectothermic and endothermic animals. We performed an experiment to test whether bones of endothermic animals, having potentially lost their ability to adapt to thermal cycling, exhibit reduced resistance to thermal fatigue compared to ectothermic animals, which may have retained this adaptive trait due to their environmental conditions. The change in stiffness were determined using shifts in the resonant peaks of the frequency spectrum obtained from Resonant Ultrasonic Spectroscopy (RUS). To achieve this, samples of compact (cortical) and spongy bone tissue were extracted and polished before undergoing a 29-day period of thermal cycling. The changes in the resonance frequencies were then observed. Changes in resonant frequencies imply corresponding changes in elastic constants. The primary findings indicated that bones from ectothermic species exhibited minimal changes in elastic properties compared to those from endothermic species, as evidenced by the smaller shifts in resonant peak magnitudes following thermal cycling.

1. **Introduction**

Emulating materials that have been naturally refined over millions of years of evolution can help overcome the current limitations of micro-architectured materials, particularly in terms of weight, fatigue and environmental resistance [3]. Recent advances in micro-fabrication techniques, like additive manufacturing with various materials such as metal alloys and ceramics, have led to the creation of innovative structures that mimic the natural properties of biological materials like bones [2, 4-6]. Bones have an exceptional specific strength due to their internal structure, making them both lightweight and durable. They also remodel themselves in response to internal or external stresses. Bone consists of two macroscopic tissues with different properties, namely trabecular and compact (cortical) bone. Trabecular bone is often found in distal regions of long bones and consists of an intricate network of osseous struts connecting regions of space, while compact bone is found on superficial layers of bone and is denser with mineralized lamellar bone arranged in the direction of routine stresses allowing it to withstand sudden impact forces [7]. With the integration of precise manufacturing techniques and microstructures inspired by biological models, it is essential to account for fatigue in these structures, as complex geometries can introduce stress concentrations at the micro-scale, potentially shortening the fatigue life of the components [8]. Biological systems are inherently self-repairing, which makes modeling those systems challenging for conventional material and manufacturing modeling techniques. Consequently, fatigue-induced stresses limit some of the advantages of biomimicry. Lightweight and strong materials are crucial for optimal performance and durability in many engineering applications, such as aerospace structures, especially under thermal fatigue conditions. Thermal fatigue plays a significant role in crack propagation, as repeated thermal stresses cause cyclic expansion and contraction. Over time, these stresses can exceed the material's elastic limit, resulting in plastic deformation and catastrophic fracture [6].

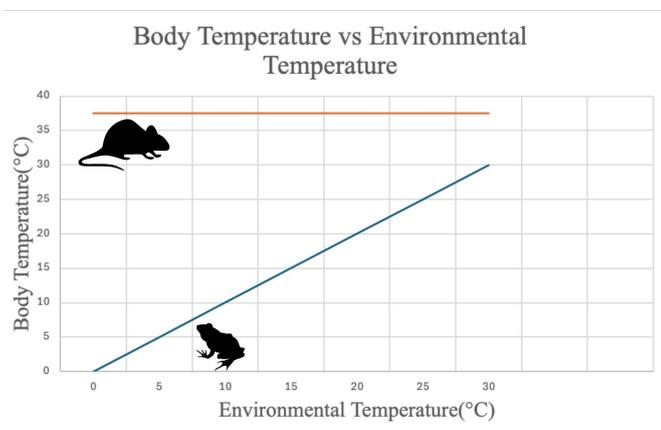

**Fig.1**: Internal temperature variation versus environmental temperature between ectotherms and endotherms[1, 2]

Investigating the differences in thermal fatigue behavior between bones of endothermic and ectothermic animals may offer valuable insights into factors contributing to thermal fatigue resistance. Bones of ectothermic vertebrates frequently experience significant temperature fluctuations, while the bones of endothermic animals remain nearly constant in temperature (see Fig. 1). This study hypothesizes that the bones of ectothermic vertebrates are adapted to withstand greater thermal stresses compared to those of endothermic animals. The bones of endothermic animals may have evolved without a need for thermal fatigue resistance. Such evolutionary differences could influence the mechanical properties and overall thermal fatigue resistance of their bones.

To compare different bones and their respective thermal fatigue resistance, resonant ultrasound spectroscopy (RUS) was employed to characterize the elastic tensor of each specimen [9-11]. Essentially, RUS measures the natural resonance frequencies of a material, providing detailed insights into its elastic properties [12-15]. Typically, the resonant response is influenced by several parameters, such as sample dimensions, elastic constants, and crystal orientation [16] and by comparing measured and calculated resonant frequencies, a specific parameter of interest can be extracted. Initially, this method was employed in geophysics to determine the Earth's internal structure from seismic data and since then it has been applied to various fields such as solid-state physics [17, 18], construction materials [19] and elastomers [20]. In the medical field, RUS has been implemented to determine the elastic constants of hydrated human dentin [21] and compact bones [22].

The most common method for determining mechanical resonances in RUS employs piezoelectric transducers to excite and measure resonant frequencies. One transducer generates a series of standing elastic waves with constant amplitude and varying frequency, while the other detects the sample's response. Resonances are identified by sweeping through the frequency range and plotting peaks corresponding to the sample's natural resonant frequencies. To derive a solid's elastic properties from its measured resonances, an "inverse computation" process is applied. This process numerically inverts the forward computation, which lacks an analytical solution. An iterative approach, such Levenberg–Marquardt method [23], is used to adjust the elastic tensor until the computed resonances align with the measured ones within a predefined error threshold. The resulting elastic tensor is considered representative of the sample's properties. In this study, the inverse computation is not performed. Instead, only the resonant frequencies are measured, without deriving the elastic constants. These resonant frequencies are functions of the elastic constants, as detailed in Section 1.1. Extracting elastic constants from resonant frequencies requires capturing at least as many resonant frequencies as there are elastic constants, with lots of additional measurements typically needed to minimize error. This aspect presented challenges in the current work. Consequently, the focus is on assessing the relative change in resonant frequencies before and after heat cycling, rather than directly determining the elastic constants.

By analyzing shifts in resonant frequencies before and after thermal cycling, this study provides insights into the thermal fatigue behavior of bone. Understanding how biological structures endure repeated thermal stresses not only advances material characterization techniques but also informs the design of engineered materials. The thermal fatigue resistance of ectothermic bones provides valuable insights for aerospace materials, which endure extreme temperature cycling in space and atmospheric re-entry. Bone's hierarchical structure, particularly its combination of lightweight strength and adaptability, serves as a model for designing fatigue-resistant materials. Mimicking these features in aerospace composites, thermal barrier coatings, or additively manufactured structures could enhance durability and performance. This study highlights the potential of bioinspired materials to improve thermal fatigue resistance in aerospace applications.

### 1.1. RUS General Overview

The deformation of a body, $u_i$ can be expressed as an infinitesimal strain tensor:

$$\epsilon_{ij} = \frac{\partial u_i}{\partial x_j} + \frac{\partial u_j}{\partial x_i} \tag{1}$$

Since the strain tensor is symmetric, the potential energy is expressed as:

$$PE = \frac{1}{2} C_{ijkl} \frac{\partial u_i}{\partial x_j} \frac{\partial u_k}{\partial x_l} \tag{2}$$

The Lagrangian for an elastic body with free surfaces is given by:

$$L = \int_V \left( \frac{1}{2} \rho \omega^2 u_i u_i - \frac{1}{2} C_{ijkl} \frac{\partial u_i}{\partial x_j} \frac{\partial u_k}{\partial x_l} \right) dV \tag{3}$$

where a sinusoidal time dependence with angular frequency ω is assumed, and the integral is over the volume of the body. The components of the displacement vector, $u_i$, that solve the elastic wave equation are determined by extremizing $L$ in Eq. (3). The Rayleigh-Ritz method is used to find this extremum by expanding the displacement vector in terms of a complete set of basis functions:

$$u_i = \sum_q a_{iq} \varphi_q \tag{4}$$

Substituting $u_i u_i$ in the Lagrangian by $\delta_{ii'} u_i u_{i'}$ the expression becomes:

$$L = \frac{1}{2} \rho \omega^2 a_{iq} a_{i'q'} \int_V \delta_{ii'} \varphi_{iq} \varphi_{i'q} \, dV - \frac{1}{2} a_{iq} a_{kq'} \int_V C_{ijkl} \frac{\partial \varphi_q}{\partial x_j} \frac{\partial \varphi_{q'}}{\partial x_l} \, dV \tag{5}$$

This can be written more compactly as:

$$L = \frac{1}{2} \rho \omega^2 a^T \mathbf{E} a - \frac{1}{2} a^T \mathbf{\Gamma} a \tag{6}$$

where **E** and **Γ** are volume integrals involving the basis functions, their derivatives, and and $C_{ijkl}$. The coefficients $a_{iq}$ are treated as vectors. To find $a$ that extremizes L, the necessary condition is:

$$\rho \omega^2 E a = \Gamma a \tag{7}$$

This represents a general eigenvalue problem. In this study, inverse RUS calculations, which derive elastic constants from resonant frequencies, were not performed. However, as demonstrated, resonant frequencies are directly influenced by the elastic constants. Thus, changes in resonant frequencies serve as indicators of corresponding changes in elastic constants.

## 2. Materials and Methods/Procedure

This study analyzed 29 specimens of compact and spongy bone tissue from three species with differing thermoregulatory mechanisms: two endothermic species, horse (equus ferus caballus) and cow (bos taurus), and one ectothermic species, American alligator (alligator mississippiensis). Two mammalian species were selected to account for interspecies variations in bone properties such as density and fracture stress [24]. A mississippiensis was chosen due to its comparable adult body mass to the selected mammals, making it suitable for comparison with mammals subjected to similar mechanical stresses. The complete list of samples is provided in Table 1.

Large reptiles, such as alligators and crocodiles, are known to exhibit inertial homeothermy, maintaining relatively stable body temperatures due to their large size and thermal inertia [25], particularly in environments with consistent temperatures. However, under fluctuating ambient temperatures, the body temperatures of large reptiles can vary substantially over periods of hours or days, despite the mitigating effects of gigantothermy [26].

Table 1 Experimental Sample Composition

|  | Species | Number of Individuals | Number of Samples |
|---|---|---|---|
| Femur | Cow | 3 | 6 compacts |
|  | Horse | 1 | 3 compacts |
|  | Alligator | 3 | 4 compacts, 2 spongy |
| Humerus | Cow | 1 | 2 compacts |
|  | Horse | 1 | 2 compacts |
|  | Alligator | 3 | 6 compacts, 4 spongy |

The sample preparation methodology was adapted primarily from the work of Kinney et al. [27]. A slow-speed diamond-tipped drill bit was identified as the most suitable tool. Humeri and femora were sectioned into approximately 1-inch-thick discs through axial cutting. These discs were secured in a bench vise (see Fig. 2) while cylindrical samples were cored using a Jet® drill press equipped with a diamond-tipped hollow drill bit (6 mm internal diameter). Cores were extracted exclusively from either the spongy or compact regions, ensuring no overlap between the two, and were consistently sourced from anatomically similar regions across all species.

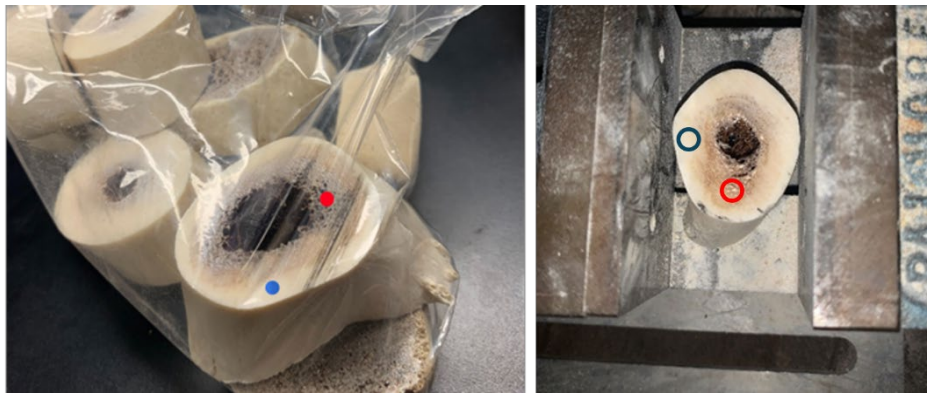

**Fig.2.** Sawed disks ready for coring; red circle indicates spongy bone and blue circle indicates compact bone

The resulting cored samples exhibited uniform diameters but variable lengths. To achieve uniform length and flat end surfaces, a rotary tool with sandpaper discs was employed at the lowest rotational speed. This process shaped the samples into comparable dimensions. Following preparation, the samples were cleaned using compressed air and stored in airtight, individual bags to prevent contamination. For all subsequent measurements and procedures, gloves and tweezers were used to handle the samples, minimizing the risk of contamination from oils or other residues.

**2.1. RUS Measurement Protocol**

Resonant Ultrasound Spectroscopy (RUS) measurements were performed using a spectrometer developed by Alamo Creek Engineering. To optimize the number of excited resonant frequencies, each sample was positioned asymmetrically and held by the edges of the cylinder, as illustrated in Fig. 3. Measurements were conducted by transmitting signals through a drive transducer, propagating through the sample, and receiving the signals with a pickup transducer. The resulting data were displayed as a spectrum (see Fig. 3 for a schematic view). RUS measurements were conducted both before and after heat cycling, the details of which are provided in Section 2.2.

RUS measurements were made on each of the 29 samples. Once a sample was placed between the two transducers of the RUS device, a configuration file was created to optimize the settings which included the start frequency, end frequency, output amplitude, coarse step, fine step, and step duration. Once optimized, the configuration file was saved and used for that specific sample to ensure consistency in the measurements before and after thermal cycling.

A common challenge in RUS measurement is that, even with optimal sample placement between the transducers, some resonant modes may not be excited and thus may be absent from the spectrum. To address this limitation, multiple measurements with varying sample orientations and placements were performed to maximize the excitation of resonant modes. As such, the following procedure was implemented: After creating the configuration file, each sample underwent edge-to-edge measurement, where the cylindrical sample was positioned such that, its edge was in contact with the piezoelectric transducers. During this process, the resonance spectrum was recorded, and a frequency plot was generated. The resulting data (magnitude in volts and frequency in hertz) were exported as CSV files and organized in a spreadsheet. For each sample, the edge-to-edge measurement was repeated after a 90-degree rotation, with the process repeated for three additional 90-degree rotations. The sample was then flipped, and the same measurements were performed on the opposite side, resulting in a total of eight datasets (magnitude vs. frequency) per sample. The collected datasets were post processed and the averages of the magnitudes were calculated to determine the representative resonant frequencies for each sample. This approach ensured comprehensive excitation of resonant modes.

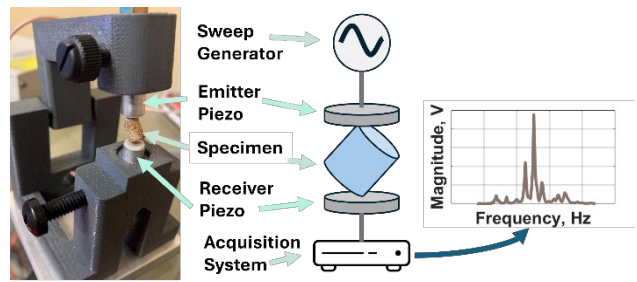

**Fig.3:** The Resonant Ultrasound Spectroscopy (RUS) measurement device and corresponding schematic illustration.

The primary aim, initially, was to measure the complete elastic constants of the bone samples. However, as noted in prior studies [28], RUS has limitations when applied to highly anisotropic, inhomogeneous bone samples with significant surface irregularity. Additionally, the limited number of measurable resonant frequencies restricted the accuracy of inverse RUS calculations. Nevertheless, the precision of resonance frequency measurements allowed shifts in resonant peaks to serve as reliable proxies for changes in sample elastic constants. Specifically, a downward shift in resonant frequencies corresponded to a reduction in elastic constants, effectively indicating changes in material properties.

The collected data were exported to a spreadsheet for preliminary organization, post processed, and subsequently imported into MATLAB for further analysis and quantification of frequency shifts. Statistical analyses were conducted using Design-Expert (a commercial software) to evaluate and characterize the factorial significance of the observed responses.

### 2.1.1. Control Measures for RUS

Several controls were implemented to ensure the reliability and accuracy of the procedure. First, duplicate edge-to-edge measurements were performed on both sides of each sample. Averaging these readings helped validate measurement consistency. Second, repeat characterization was conducted on one randomly selected sample from each bone of each species, both before and after thermal cycling, to confirm that any observed excitation frequency shifts were attributable to thermal stress. Third, six samples (two per species) served as non-cycled controls. These samples were prepared and measured using the same protocol but were not subjected to thermal cycling. To further ensure accuracy, the non-cycled control samples were measured 30 days apart to confirm that frequency shifts in thermally cycled samples were due to thermal cycling rather than material relaxation.

This comprehensive protocol facilitated robust data collection and enabled precise analysis of resonance behavior under different thermal conditions.

### 2.2. Thermal Cycling Protocol

For the thermal cycling conducted in this study, a RapidTherm® thermal chamber equipped with an IDEC touchscreen controller was utilized. The IDEC controller employs a cascaded control strategy, incorporating two temperature channels and two temperature sensor inputs, to achieve rapid temperature changes and precise temperature control for the device under test (DUT). The controller features a "thermal boost" mechanism that dynamically adjusts the air temperature supplied to the chamber's workspace, either overdriving or underdriving it to match the target temperature. Air temperature regulation is managed by setpoint 2, while setpoint 1 represents the user-defined temperature for the DUT.

A sophisticated fan system enhances heat transfer throughout the chamber by automatically adjusting the speed of the fan motor(s) via a variable frequency drive. This ensures optimal airflow and facilitates rapid temperature adjustments to the DUT in response to changing setpoints, enabling efficient and precise thermal cycling.

For the experiment, samples were affixed to two-sided tape to prevent displacement due to airflow and were arranged in a predetermined grid on the chamber floor. Before cycling began, the samples were equilibrated at room temperature (25°C) for 1 minute. The internal temperature was then reduced to 0°C over 5 minutes at a rate of 5°C per minute. Once at 0°C, the samples were held at this temperature for 10 minutes to ensure complete material contraction. The temperature was subsequently increased to 40°C over 4 minutes at a rate of 10°C per minute. Following this, the samples were held at 40°C for 10 minutes to promote material expansion[29, 30].

This cycle was repeated 1,400 times over approximately 29 days, as detailed in Table 1. After the final cycle, the thermal cycling process was terminated. This structured protocol was designed to simulate the thermal variations that large alligators, may experience in their natural environments. The procedure provided a rigorous test of thermal cycling resistance and material behavior under fluctuating temperatures, consistent with the hypothesized adaptive temperature range for these species [31].

**Table 2: Thermal Cycling Protocol**

| Step Number | Step Type | Step Time (min) | Setpoint (°C) | Thermal Boost | Deg/min |
|---|---|---|---|---|---|
| 1 | Ramp Time | 1 | 25 | 10 | 0 |
| 2 | Ramp Time | 5 | 0 | 10 | 5 |
| 3 | Soak | 10 | N/A | N/A | N/A |
| 4 | Ramp Time | 4 | 40 | N/A | 10 |
| 5 | Soak | 10 | N/A | N/A | N/A |
| 6 | Loop | N/A | N/A | N/A | N/A |
| 7 | End of Loop | N/A | N/A | N/A | N/A |

## 3. Results and Discussion

The primary objective of the RUS data analysis was to quantify potential differences in fatigue resistance between bones from endothermic and ectothermic species. Across all 29 samples, the resonant frequency shifts in bones from endothermic species were consistently larger than that observed in an ectothermic specie, as illustrated by representative examples in Fig. 4.

### 3.1. Statistical Analysis

A systematic protocol was developed to quantify the observed shifts in resonant frequencies. This process involved analyzing pre-thermal and post-thermal resonance spectra using a MATLAB script. The script processed input files containing frequency data, identified peaks of interest, and prompted the user to select three peaks. For each peak, horizontal line segments were drawn at two-thirds of the peak height (Fig. 4, top graph), and the midpoint of these segments was calculated. Differences in midpoints between pre- and post-thermal cycling plots were computed for the three peaks, and their average was used as a quantifier of overall frequency shift.

This average shift served as an indicator of changes in the elastic properties of the samples. Bones from endothermic species exhibited significant shifts in resonant frequencies post-thermal cycling, regardless of bone type (cortical or trabecular). In contrast, shifts in ectothermic samples were minimal.

#### 3.1.1. Design of Experiments and Response Variables

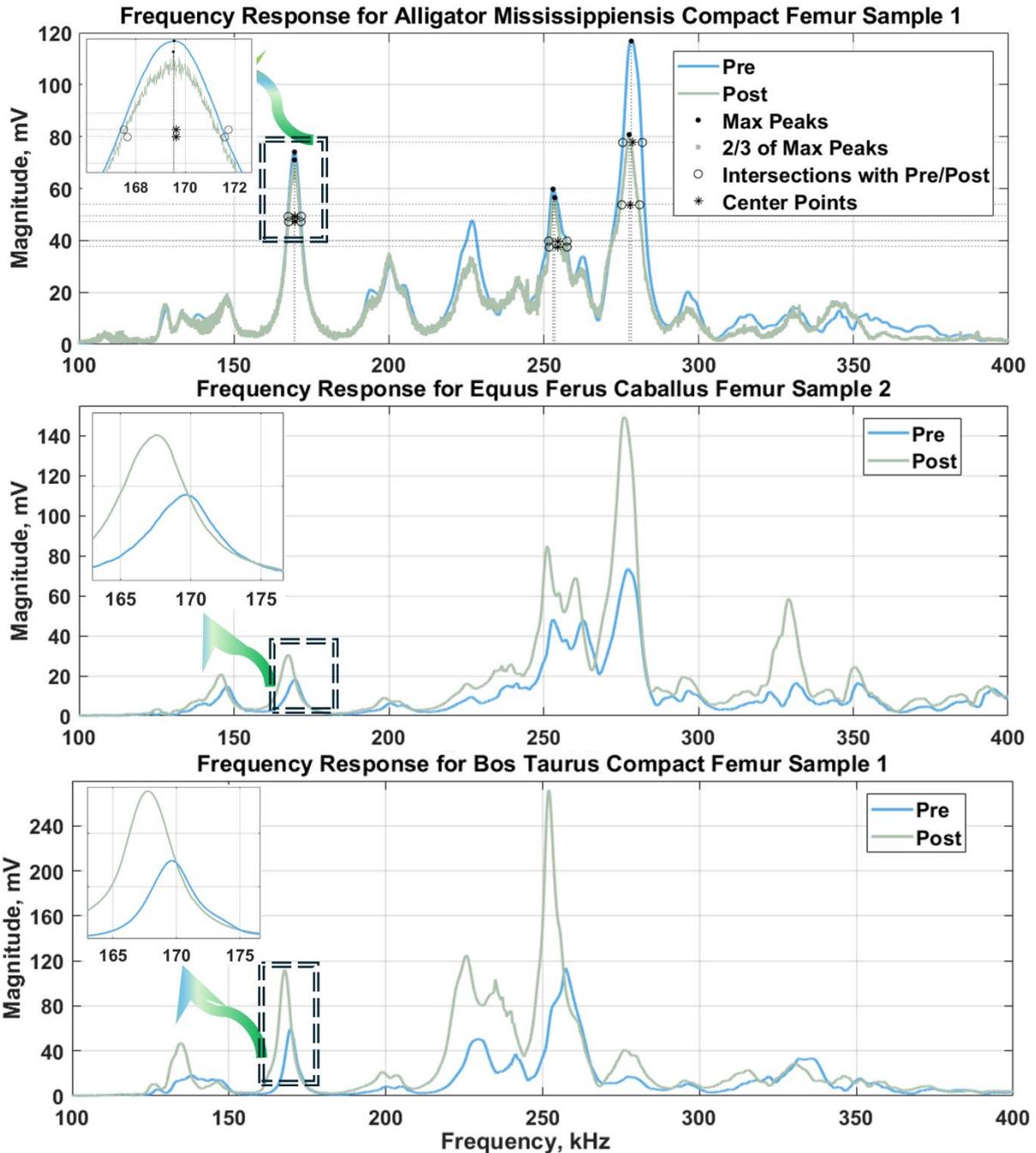

**Fig 4:** Resonant frequency shifts observed in compact bone samples from three species: (Top) Alligator mississippiensis, showing minimal changes between pre- and post-thermal cycling; (Middle) Equus ferus caballus, displaying generally greater frequency shifts compared to ectothermic samples, though some shifts are similar to those of ectotherms; (Bottom) Bos taurus, exhibiting frequency shifts consistent with the pattern observed in Equus ferus caballus and notable changes between pre- and post-thermal cycling.

Statistical analyses were conducted using Design-Expert commercial software to evaluate the effects of experimental factors on the observed frequency shifts. The study was modeled as a three-factor, three-response design:

- Factor A: Bone position (femur or humerus),
- Factor B: Species (endothermic or ectothermic),
- Factor C: Bone composition (compact or spongy bone).

Three response variables were defined:

- Response 1: Shift – The average frequency shift calculated using the MATLAB protocol.
- Response 2: Absolute Individual Shift – The absolute values of the shifts for the three peaks, averaged. This was introduced to account for alternating positive and negative shifts observed in ectothermic samples.
- Response 3: Area Under the Curve (AUC) – A metric representing the area under the resonance spectrum curve, included to quantify changes in resonance magnitude, which were more pronounced in endothermic samples.

Table 3 summarizes the factors, responses, and values. The results showed that for endothermic species, Responses 1 and 2 were nearly identical, as frequency shifts were predominantly in the same direction. For ectothermic species, however, the differences between these responses were significant due to alternating positive and negative shifts. AUC shifts were consistently more substantial in endothermic samples.

**Table 3:** Input data used in Design-Expert software for evaluating the effects of input factors on measured responses.

| Standard Order | Run Order | Factor A | Factor B | Factor C | Response 1 | Response 2 | Response 3 |
|---|---|---|---|---|---|---|---|
| 1 | 1 | Femur | Cow | Compact | -3427 | 3427 | 3540 |
| 13 | 2 | Femur | Cow | Compact | -5959 | 5959 | 1450 |
| 14 | 3 | Femur | Cow | Compact | -2617 | 2617 | 1060 |
| 15 | 4 | Femur | Cow | Compact | -1881 | 1881 | 4480 |
| 16 | 5 | Femur | Cow | Compact | -2646 | 2646 | 1680 |
| 17 | 6 | Femur | Cow | Compact | -1411 | 1411 | 2190 |
| 2 | 7 | Humerus | Cow | Compact | -1117 | 1528 | 2230 |
| 12 | 8 | Humerus | Cow | Compact | -1323 | 1323 | 1210 |
| 3 | 9 | Femur | Alligator | Compact | -265 | 340 | -349 |
| 18 | 10 | Femur | Alligator | Compact | -59 | 1764 | 325 |
| 19 | 11 | Femur | Alligator | Compact | 0 | 176.333 | 1480 |
| 20 | 12 | Femur | Alligator | Compact | 120.666 | 709.3 | 1160 |
| 7 | 13 | Femur | Alligator | Spongey | -50 | 369 | -959 |
| 21 | 14 | Femur | Alligator | Spongey | -353 | 646 | 300 |
| 8 | 15 | Humerus | Alligator | Spongey | -67 | 874 | -920 |
| 27 | 16 | Humerus | Alligator | Spongey | -370 | 839.666 | 190 |
| 28 | 17 | Humerus | Alligator | Spongey | 0 | 671.666 | 190 |
| 29 | 18 | Humerus | Alligator | Spongey | -403 | 470.333 | 341 |
| 4 | 19 | Humerus | Alligator | Compact | -75 | 378 | 1040 |
| 22 | 20 | Humerus | Alligator | Compact | -1399 | 1398.5 | 720 |
| 23 | 21 | Humerus | Alligator | Compact | -1172 | 1247.5 | -750 |
| 24 | 22 | Humerus | Alligator | Compact | -101 | 134.666 | -1430 |
| 25 | 23 | Humerus | Alligator | Compact | 617 | 617 | 452 |
| 26 | 24 | Humerus | Alligator | Compact | 29 | 147.333 | 2190 |
| 6 | 25 | Humerus | Horse | Compact | -2999 | 2999 | 2300 |
| 9 | 26 | Humerus | Horse | Compact | -2318 | 2318 | 5560 |
| 5 | 27 | Femur | Horse | Compact | -1713 | 1713 | -307 |
| 10 | 28 | Femur | Horse | Compact | -1705 | 1705 | 2170 |
| 11 | 29 | Femur | Horse | Compact | -2969 | 2969 | -717 |

### 3.1.2. ANOVA and Factor Analysis

Analysis of variance (ANOVA) was performed using Design-Expert to determine the significance of the factors influencing the responses. The F-value assessed whether the variance explained by a factor was significantly greater than random error, while the p-value determined the statistical significance of the F-value. A p-value < 0.05 indicated a significant effect.

As shown in Table 4, Factor B (Species) was the most significant determinant across all three responses. The lack-of-fit F-values were not significant relative to the pure error, confirming the robustness of the model. Additionally, Shapiro-Wilk tests validated the statistical significance of the findings.

**Table 4**: ANOVA for Factor B's Effect on Responses

| Response | Source | Sum of Squares | Degrees of Freedom | Mean Square | F-Value | P-Value | Significance |
|---|---|---|---|---|---|---|---|
| **1: Shift** | **Model** | 3.632×10⁷ | 2 | 1.816×10⁷ | 20.72 | <0.0001 | significant |
| | **B-Species** | 3.632×10⁷ | 2 | 1.816×10⁷ | 20.72 | <0.0001 | |
| | **Residual** | 2.280×10⁷ | 26 | 8.768×10⁵ | | | |
| | Lack of Fit | 5.254×10⁶ | 5 | 1.051×10⁶ | 1.26 | 0.3185 | not significant |
| | Pure Error | 1.754×10⁶ | 21 | 8.353×10⁵ | | | |
| | **Cor Total** | 5.912×10⁷ | 28 | | | | |
| **2: Absolute Individual Shift** | **Model** | 2.411×10⁷ | 2 | 1.206×10⁷ | 14.53 | <0.0001 | significant |
| | **B-Species** | 2.411×10⁷ | 2 | 1.206×10⁷ | 14.53 | <0.0001 | |
| | **Residual** | 2.157×10⁷ | 26 | 8.296×10⁵ | | | |
| | Lack of Fit | 4.095×10⁶ | 5 | 8.189×10⁵ | 0.984 | 0.0923 | not significant |
| | Pure Error | 1.748×10⁷ | 21 | 8.322×10⁵ | | | |
| | **Cor Total** | 4.568×10⁷ | 28 | | | | |
| **3: Delta Area Under Curve** | **Model** | 2.344×10⁷ | 2 | 1.172×10⁷ | 5.93 | 0.0078 | significant |
| | **B-Species** | 2.344×10⁷ | 2 | 1.172×10⁷ | 5.93 | 0.0078 | |
| | **Residual** | 4.942×10⁷ | 25 | 1.977×10⁶ | | | |
| | Lack of Fit | 1.764×10⁷ | 5 | 3.529×10⁶ | 2.22 | 0.0923 | not significant |
| | Pure Error | 3.178×10⁷ | 20 | 1.589×10⁶ | | | |
| | **Cor Total** | 7.286×10⁷ | 27 | | | | |

## 4. Conclusion and Future Work

This study investigated the effects of thermal cycling on the elastic properties of bones from endothermic and ectothermic species. RUS analysis revealed distinct differences in the responses of the two groups. Bones from ectothermic species exhibited minimal shifts in resonant frequencies, suggesting high thermal fatigue resistance. Conversely, bones from endothermic species showed significant frequency shifts, indicating notable changes in elastic properties after thermal cycling.

These differences may arise from structural or compositional variations at the micro- or nanoscale, or a combination of factors. Further investigation is required to elucidate the mechanisms underlying these effects.

The insights from this research have potential applications in material science, particularly in the design of aerospace components. By mimicking the thermal fatigue resistance of ectothermic bones, materials can be developed for use in aircraft and spacecraft components subjected to extreme cyclic temperature variations. Such advancements could lead to safer, more durable materials for aerospace and propulsion systems, enhancing performance in demanding environments.